\documentclass[a4paper,aps,prd,10pt,preprintnumbers,showpacs,twocolumn,superscriptaddress,nofootinbib,amsmath,amssymb,amsfonts]{revtex4-1}
\usepackage{cmap,graphicx}
\usepackage[utf8]{inputenc}
\usepackage[T1]{fontenc}

\def\imo{i}
\def\Order#1{\mathcal{O}\left(#1\right)}

\begin{document}

\title{Nonoscillatory gravitational quasinormal modes\\of Reissner-Nordström-de Sitter spacetime}

\author{Z. Stuchlík}\email{zdenek.stuchlik@physics.slu.cz}
\affiliation{Research Centre for Theoretical Physics and Astrophysics, Institute of Physics, Silesian University in Opava, Bezručovo náměstí 13, CZ-74601 Opava, Czech Republic}

\author{A. Zhidenko}\email{olexandr.zhydenko@ufabc.edu.br}
\affiliation{Research Centre for Theoretical Physics and Astrophysics, Institute of Physics, Silesian University in Opava, Bezručovo náměstí 13, CZ-74601 Opava, Czech Republic}
\affiliation{Centro de Matemática, Computação e Cognição (CMCC), Universidade Federal do ABC (UFABC),\\ Rua Abolição, CEP: 09210-180, Santo André, SP, Brazil}


\begin{abstract}
We revisit quasinormal modes of Reissner-Nordström-de Sitter black holes, a classical and seemingly well-understood solution of Einstein-Maxwell theory with a positive cosmological constant. While previous studies have focused almost exclusively on the oscillatory photon-sphere branch, we show that two additional branches of nonoscillatory, purely imaginary modes -- associated with the de Sitter horizon and with the regime of near-extremal charge (zero-damped) -- are essential for a complete description of the spectrum. These modes have so far remained largely unexplored for gravitational perturbations despite their relevance for late-time dynamics and stability. In particular, we derive accurate analytic expressions for the purely imaginary quasinormal frequencies in the slow-damping regime, providing a complete description of this sector of the spectrum.
\end{abstract}

\maketitle

\section{Introduction}

Quasinormal modes (QNMs) \cite{Kokkotas:1999bd,Berti:2009kk,Nollert:1999ji,Konoplya:2011qq,Bolokhov:2025uxz} represent fundamental oscillatory characteristics of black hole geometries, emerging naturally as resonant frequencies of perturbations in the spacetime surrounding black holes. These modes are directly observable through gravitational wave signals detected by observatories such as LIGO and Virgo \cite{LIGOScientific:2016aoc,LIGOScientific:2017vwq,LIGOScientific:2020zkf}, providing a powerful tool for probing the properties of black holes and testing general relativity in the strong-field regime. Consequently, QNMs of a wide array of black hole models, including both classical and exotic configurations, have been extensively investigated across an enormous number of papers.

Considerable literature is devoted to quasinormal spectra of black holes in various large scale cosmological environments \cite{Konoplya:2021ube,Konoplya:2005sy,Jusufi:2019ltj,Zhang:2021bdr,Hamil:2025pte,Lutfuoglu:2025hjy,Lutfuoglu:2025kqp}.
Particularly, the spectra of quasinormal modes in asymptotically de Sitter spacetimes have attracted significant attention \cite{Zhidenko:2003wq,Yoshida:2003zz,Konoplya:2004uk,Cardoso:2003sw,Tattersall:2018axd,Aragon:2020tvq,Hintz:2021vfl,Churilova:2021nnc,Roussev:2022hpm,Konoplya:2007zx,Dyatlov:2011jd,Konoplya:2003dd,Molina:2003ff,Fernando:2015fha,Natario:2004jd,Konoplya:2008au,Konoplya:2017ymp,Toshmatov:2017qrq,Cuyubamba:2016cug,Konoplya:2013sba,Konoplya:2025mvj,Dubinsky:2024gwo,Dubinsky:2024hmn,Konoplya:2023moy,Churilova:2021nnc,Konoplya:2020fwg,Kanti:2005xa,Malik:2024bmp}. These studies are not only motivated by the growing evidence for a positive cosmological constant from observational cosmology but also by the distinctive features of the quasinormal spectrum due to existence of the de Sitter horizon. The interest in QNMs has been further stimulated by the proposed implications of the Strong Cosmic Censorship limit \cite{Cardoso:2017soq,Dias:2018ynt,Dias:2018ufh,Mo:2018nnu,Liu:2019lon,Hod:2020ktb}.

A peculiar aspect of perturbations in asymptotically de Sitter spacetimes lies in their late-time behavior. Unlike asymptotically flat spacetimes, where perturbations typically decay following a power-law tail \cite{Price:1971fb}, in de Sitter spacetimes \emph{exponential tails} prevail at late times (except when the black hole size is comparable to the cosmological scale). This behavior was first demonstrated for scalar field perturbations in \cite{Brady:1996za} and later for gravitational perturbations and fields of nonzero spin in \cite{Brady:1999wd}. However, accurately identifying the asymptotic tail behavior from numerical simulations remains a challenging task due to the intricate dependence of the decay law on various physical parameters \cite{Molina:2003dc}.

The nature of these exponential tails was revealed in \cite{Konoplya:2022xid,Konoplya:2024ptj}, where it was shown that the exponential tails of the Schwarzschild-de Sitter black hole represent a distinct phase of the quasinormal ringing, governed by purely imaginary quasinormal modes. Unlike the algebraically special mode, which satisfies total transmission boundary conditions \cite{MaassenvandenBrink:2000iwh,Cook:2016ngj}, these purely imaginary modes satisfy the ingoing boundary condition at the event horizon and thus indeed belong to the quasinormal spectrum.

Notably, in the limit of vanishing black hole mass, these modes continuously transition into the modes of empty de Sitter spacetime \cite{Lopez-Ortega:2006tjo,Lopez-Ortega:2006aal,Jansen:2017oag}. In the limit of a vanishing cosmological constant, the purely imaginary quasinormal frequencies asymptotically approach zero, indicating that these modes dominate the spectrum with the lowest damping rate when the cosmological constant is small, suggested by observational cosmology. It is also worth mentioning that this branch of de Sitter modes breaks down the correspondence between null geodesics and eikonal quasinormal modes \cite{Konoplya:2022gjp}.

For the Reissner-Nordström-de Sitter spacetime, purely imaginary quasinormal modes were previously identified for a test scalar field \cite{Jansen:2017oag,Cardoso:2017soq} and the corresponding pseudospectrum was analyzed in \cite{Sarkar:2023rhp}. However, to the best of our knowledge, a comprehensive analysis of such modes within the gravitational sector, which also includes the electromagnetic perturbations, has not yet been conducted.

Another branch of quasinormal modes, characterized by purely imaginary frequencies for the Reissner-Nordström-de Sitter spacetime, emerges in the near-extremal regime of black holes. This branch was first identified for Kerr black holes in Refs.~\cite{Detweiler:1980gk,Glampedakis:2001js,Hod:2008zz} and was later called zero-damped modes (ZDMs), reflecting the fact that their imaginary parts vanish in the limit of extremal rotation \cite{Yang:2012pj,Richartz:2017qep}. Subsequently, analogous modes were discovered in charged black hole spacetimes, including Reissner-Nordström \cite{Kim:2012mh} and Kerr-Newman black holes \cite{Konoplya:2013rxa,Mark:2014aja}. For a massive, charged scalar field in the Reissner-Nordström spacetime, Ref.~\cite{Hod:2017gvn} derived an analytic expression for ZDMs characterized by damping rates that vanish as the black hole charge approaches its extremal value. For massless and uncharged fields these modes are purely imaginary. In the case of gravitational perturbations of the Reissner-Nordström black hole, ZDMs were first computed in Ref.~\cite{Zimmerman:2015trm}. However, the ZDMs in the presence of a cosmological constant were discussed only for a test scalar field in the context of the Strong Cosmic Censorship conjecture \cite{Cardoso:2017soq}.

In this work, we address this gap by calculating the purely imaginary quasinormal modes of gravitational perturbations of the Reissner-Nordström-de Sitter black hole. We employ the Leaver method to determine these modes in the frequency domain. Furthermore, by integrating the wave equation in the time domain, we demonstrate that the observed exponential tails are, in fact, a manifestation of these purely imaginary quasinormal modes.

Our paper is organized as follows: In Sec.~\ref{sec:formulas} we give the basic formulas on the black hole metric and the form of the wave equation describing coupled gravitational and electromagnetic perturbations. Sec.~\ref{sec:numericalmethods} is devoted to the brief description of the numerical methods we use: Leaver method and time-domain integration. Sec.~\ref{sec:qnms} discusses the quasinormal spectrum of the Reissner-Nordström-de Sitter spacetime: Particularly, in Sec.~\ref{sec:complex} we review the oscillatory quasinormal modes, in Sec.~\ref{sec:zdms} we study the zero-damped modes of near-extremally charged black holes, in Sec.~\ref{sec:deSitter} we discuss the features of the de Sitter branch of the quasinormal spectrum, associated with perturbations of de Sitter universe and in Sec.~\ref{sec:dominantmodes} its interplay with the branch of oscillatory modes, associated with the black hole. Finally, in the Conclusions we summarize the obtained results.

\section{Gravitational perturbations of the Reissner-Nordström-de Sitter black holes}\label{sec:formulas}

Here we will use the convenient formalism for $4$-dimensional Reissner-Nordström-de-Sitter black holes and designations of \cite{Kodama:2003jz,Ishibashi:2003ap,Kodama:2003kk}, taking then $D=n+2=4$. The metric has the following form:
\begin{equation}\label{metric}
ds^2=-f(r)dt^2+\frac{dr^2}{f(r)}+r^2(d\theta^2+sin^2\theta
d\phi^2).
\end{equation}
where
\begin{eqnarray}\label{metric-function}
&f(r)=1-X+Z-Y,&
\\\nonumber
&X=\dfrac{2M}{r},\qquad Y=\dfrac{\Lambda r^2}{3}, \qquad Z=\dfrac{Q^2}{r^{2}}.&
\end{eqnarray}

The perturbation equations can be reduced to the wavelike form \cite{Kodama:2003kk}
\begin{equation}\label{hyperbolic}
\left(\frac{\partial^2}{\partial t^2}-\frac{\partial^2}{\partial r_*^2}\right)\Psi(t,r)=-V(r)\Psi(t,r),
\end{equation}
where \emph{the tortoise coordinate} $r_*$ is defined as
\begin{equation}\label{tortoise}
dr_*=\frac{dr}{f(r)}.
\end{equation}

The effective potentials for the decoupled vector (axial) gravitational and electromagnetic perturbations are
\begin{equation}\label{potential-vectortype}
V_{V\pm}(r)=\frac{f(r)}{r^2}\left(\lambda+2+\frac{8 Z- 3 X\pm\sqrt{9X^2+16 Z\lambda}}{2}\right),
\end{equation}
where
$$\lambda\equiv(\ell+2)(\ell-1)>0,$$
and $\ell=2,3,4,\ldots$ is the multipole number.

We note that, since $X^2/Z\equiv4M^2/Q^2$ is a constant, the effective potential (\ref{potential-vectortype}) is a rational function of the radial coordinate $r$.

The effective potentials for the decoupled perturbations of scalar (polar) type are
\begin{equation}\label{potential-scalartype}
V_{S\pm}(r)=f(r)\frac{\widetilde{U}_{\pm}(r)}{r^2H_{\pm}(r)^2},
\end{equation}
where
$$
\begin{array}{rcl}
H_+(r)&=&1-3X\delta,\\
H_-(r)&=&\lambda+3X(1+\delta\lambda),\\
\widetilde{U}_+(r)&=&-162X^4\delta^3(1+\delta\lambda)+36X^3\delta^2(3\delta\lambda+4)
\\&&-36X^2\delta(2Y\delta+1)-12X\delta\lambda+4\lambda+8,\\
\widetilde{U}_-(r)&=&-162X^4\delta(1+\delta\lambda)^3-36 X^3(1+\delta\lambda)^2(3\delta\lambda-1)
\\&&-36X^2(1+\delta\lambda)(2Y(1+\delta\lambda)-\lambda)
\\&&+12X\lambda^2(1+\delta\lambda)+4\lambda^3+8\lambda^2,
\end{array}
$$
and the constant $\delta$ is defined as follows:
$$\delta\equiv\frac{1}{2\lambda}\left(\sqrt{1+\frac{4\lambda Q^2}{9M^2}}-1\right)=\frac{1}{2\lambda}\left(\sqrt{1+\frac{16Z\lambda}{9X^2}}-1\right).$$

In the limit of zero charge $Q=0$ ($Z=0$, $\delta=0$), the potentials $V_{V+}$ and $V_{S+}$ reduce to the potential of a test Maxwell field, and the potentials $V_{V-}$ and $V_{S-}$ correspond to the axial and polar gravitational perturbations, respectively, in the Schwarzschild-de Sitter background.

For $\ell=1$, only the perturbations of the Maxwell field are dynamical (see Sec.~4.2.2~of~\cite{Kodama:2003kk} for details). These perturbations are governed by the effective potential
\begin{equation}\label{l1Maxwell}
V_{\ell=1}(r)=\frac{2f(r)}{r^2}.
\end{equation}
In four-dimensional spacetime, axial and polar perturbations are governed by equations related through a Darboux transformation, meaning that the solution for one can be derived from the other. As a result, the quasinormal spectra of $V_{V+}$ and $V_{S+}$, as well as those of $V_{V-}$ and $V_{S-}$, are identical~\cite{Molina:2003dc}. From this point forward, we will refer to these modes as ``$+$'' and ``$-$'', respectively.

To simplify the analysis, we introduce the following geometric quantities: the radius of the event horizon $r_h$, the cosmological horizon $r_c$, and the inner horizon $r_i$ \cite{Stuchlik:2002tj}. We can easily express the black hole mass $M$, the cosmological constant $\Lambda$, and the electric charge $Q$ in terms of
\begin{equation}\label{bounds}
0\leq r_i\leq r_h<r_c\leq\infty.
\end{equation}
First, using that the metric function vanishes at the event horizon, we find that
\begin{equation}\label{Massdef}
2M=r_h+\frac{Q^2}{r_h}-\frac{\Lambda r_h^3}{3}.
\end{equation}
Using the similar condition for the cosmological horizon we express the $\Lambda$-term,
\begin{equation}\label{Lambdadef}
\frac{\Lambda}{3}=\frac{r_c r_h-Q^2}{r_cr_h\left(r_c^2+r_cr_h+r_h^2\right)}.
\end{equation}
Finally, we express the value of $Q$ in terms of the inner horizon,
\begin{equation}\label{Qdef}
Q^2=\frac{r_cr_hr_i(r_c+r_h+r_i)}{r_c^2+r_h^2+r_i^2+r_cr_h+r_cr_i+r_hr_i}.
\end{equation}

The uncharged black hole corresponds to $r_i=0$, so that $Q=0$. When $r_i=r_h$, we have the extremely charged black hole, $f'(r_h)=0$. The limit $r_c\to\infty$ yields the asymptotically flat Reissner-Nordström black hole solution with $\Lambda=0$ and $Q^2=r_hr_i$.

\section{Numerical methods}\label{sec:numericalmethods}

In this section, we summarize the methods used to calculate quasinormal frequencies. We employ the Leaver method \cite{Leaver:1985ax} in the frequency domain and verify the obtained results in the time domain using the Gundlach-Price-Pullin integration scheme \cite{Gundlach:1993tp}.

\subsection{Leaver method}

In order to switch to the frequency domain, we substitute
\begin{equation}
\Psi(t,r)=e^{-\imo\omega t}\Psi(r)
\end{equation}
into the Eq.~(\ref{hyperbolic}) and obtain for the radial function
\begin{equation}\label{wavelike}
\frac{d^2\Psi}{d r_*^2}+\left(\omega^2-V(r)\right)\Psi(r)=0.
\end{equation}

Written in terms of the radial coordinate, this equation takes the form
\begin{equation}
f^2(r)\frac{d^2\Psi}{dr^2}+f(r)f'(r)\frac{d\Psi}{dr}+\left(\omega^2-V(r)\right)\Psi=0.
\end{equation}
Using the geometric quantities, the metric function $f(r)$ can be factorized as
\begin{equation}
f(r)=-\frac{\Lambda}{3r^2}(r-r_i)(r-r_h)(r-r_c)(r-r_o),
\end{equation}
where
$$r_o=-(r_i+r_h+r_c).$$

Thus, for the vector-type potentials~(\ref{potential-vectortype}), Eq.~(\ref{wavelike}) is Fuchsian with five finite regular singular points,
$$r=r_i,\qquad r=r_h,\qquad r=r_c,\qquad r=r_o,\qquad r=0.$$

It is useful to note that the point $r=\infty$ is not a singular point of the radial equation for $\Lambda>0$. Indeed, introducing the compactifying variable $z=1/r$, one finds that the equation is regular at $z=0$. The finite point $r=0$ is a regular singular point. However, for the vector-type potentials this singularity is apparent and can be removed by a suitable redefinition of the radial function, similarly to the removable
singularity in the Kerr-de Sitter Teukolsky equation \cite{Hatsuda:2020sbn}. After this redefinition, the equation has four nonapparent regular singular points and can be mapped by a fractional-linear transformation to the standard Heun equation.

An important limiting case is the asymptotically flat limit $\Lambda\to0$, in which the cosmological horizon is pushed to infinity.
In this case
\begin{equation}
f(r)=\left(1-\frac{r_i}{r}\right)\left(1-\frac{r_h}{r}\right).
\end{equation}
After the compactification $z=1/r$, one finds that $z=0$, corresponding to $r=\infty$, becomes an irregular singular point.

In the extremal-charge limit, $r_i=r_h$, two regular singular points merge, producing a double zero of $f(r)$. For nonzero frequency, the corresponding degenerate horizon becomes an irregular singular point, and this case must be treated separately \cite{Richartz:2015saa}. Similarly, the coalescence $r_h=r_c$ may be understood as a confluent limit in which the event and cosmological horizons merge and the degenerate horizon becomes an irregular singular point. However, the quasinormal ringing considered in this work is defined and observed in the static region $r_h<r<r_c$. Therefore, the direct Nariai limit is not part of the ringdown problem studied here, since this region collapses in the original radial coordinate. A nontrivial Nariai geometry can be obtained only after an additional near-horizon scaling of the coordinates, leading to a different limiting spacetime with its own perturbation problem \cite{Zhou:2025xta,Nakamoto:2026lyo}. This separate problem is beyond the scope of the present work. Thus, in this manuscript we do not study the limiting case of degenerate horizons, although we can approach it numerically arbitrarily closely.

For the scalar-type potentials~(\ref{potential-scalartype}), the zeros of $H_\pm(r)$ introduce additional apparent singularities in Eq.~(\ref{wavelike}). These singularities do not correspond to physical boundaries and do not affect the quasinormal-mode boundary conditions used in our analysis.

Now quasinormal frequencies can be defined as wave equation's eigenvalues $\omega_{n}$, where $n$ is the overtones number, which correspond specific boundary conditions: purely outgoing waves at the cosmological horizon $r=r_c$ ($r_*\to+\infty$) and purely ingoing waves at the black hole event horizon $r=r_h$ ($r_*\to-\infty$).
\begin{equation}
\begin{array}{rclr}
\Psi(t,r)&\propto& e^{-\imo\omega(t+r_*)},& r_*\to-\infty,\\
\Psi(t,r)&\propto& e^{-\imo\omega(t-r_*)},& r_*\to+\infty,
\end{array}
\end{equation}
yielding the following conditions for $\Psi(r)$ at the regular singular points of Eq.~(\ref{wavelike})
\begin{equation}\label{qnmbc}
\begin{array}{rclr}
\Psi(r)&\propto&(r-r_h)^{-\imo\omega/f'(r_h)},&r\to r_h+0,\\
\Psi(r)&\propto&(r_c-r)^{+\imo\omega/f'(r_c)},&r\to r_c-0.
\end{array}
\end{equation}

We study the solution in the interval $r_h<r<r_c$. Therefore, we introduce the new function, $y(r)$, such that
\begin{equation}\label{reg}
\Psi(r)=\left(\frac{r_c-r}{r-r_i}\right)^{\imo\omega/f'(r_c)}\left(\frac{r-r_h}{r-r_i}\right)^{-\imo\omega/f'(r_h)}y(r).
\end{equation}
Here $\Psi(r)$ satisfies the quasinormal boundary conditions (\ref{qnmbc}) if and only if $y(r)$ is regular at $r=r_h$ and $r=r_c$. We express $y(r)$ in terms of the Frobenius series expansion:
\begin{equation}\label{Frobenius}
y(r)=\sum_{k=0}^{\infty}a_k\left(\frac{r-r_h}{r-r_i}\cdot\frac{r_c-r_i}{r_c-r_h}\right)^k.
\end{equation}
By accounting for the singular point $r=r_i$ of Eq.~(\ref{wavelike}), we ensure the convergence of the series (\ref{Frobenius}) for $r_h\leq r<r_c$.
The coefficients $a_i$ satisfy a seven-term recurrence relation for vector-type perturbations (\ref{potential-vectortype}) and a nine-term recurrence relation for scalar-type perturbations (\ref{potential-scalartype}). These recurrence relations can be numerically reduced to a three-term relation via Gaussian elimination (for details, see \cite{Konoplya:2011qq}). The final step involves solving an infinite continued fraction equation for $\omega$, which ensures the convergence of the series~(\ref{Frobenius}) at $r=r_c$.

In order to improve convergence of the infinite continued fraction for the purely imaginary modes we use the Nollert method \cite{Nollert:1993zz,Zhidenko:2006rs}.
The continued-fraction approach has been discussed and used for finding precise values of quasinormal frequencies in numerous works (see \cite{Konoplya:2023ahd,Konoplya:2022iyn,Zinhailo:2024jzt,Bolokhov:2023bwm,Bolokhov:2023ruj,Konoplya:2024lch} for recent examples) and, therefore, we will not review it here in more detail.

\subsection{Time-domain integration}

In order to analyze the perturbation evolution at all times and check the values of the quasinormal modes obtained in frequency domain, we use the time-domain integration method. This approach involves solving the master equation (\ref{hyperbolic}) in terms of the light-cone coordinates, defined as $u\equiv t-r_*$ and $v\equiv t+r_*$.

For numerical integration, we utilize the discretization scheme, proposed in~\cite{Gundlach:1993tp},
\begin{eqnarray}\label{Discretization}
\Psi\left(N\right)&=&\Psi\left(W\right)+\Psi\left(E\right)-\Psi\left(S\right) \\ \nonumber&&
-\Delta^2V\left(S\right)\frac{\Psi\left(W\right)+\Psi\left(E\right)}{8}+{\cal O}\left(\Delta^4\right).
\end{eqnarray}
In the above relation, the following notations for the points were used:
$N\equiv\left(u+\Delta,v+\Delta\right)$, $W\equiv\left(u+\Delta,v\right)$, $E\equiv\left(u,v+\Delta\right)$, and $S\equiv\left(u,v\right)$.
Then, with the Gaussian initial data on the surfaces, $u=u_0$ and $v=v_0$, we calculate the values of the functions in the rhombus $u_0<u<u_1$ and $v_0<v<v_1$.
The dominant modes can be extracted from the time-domain profiles with the help of the Prony method, which is based on the fitting of the time-domain profile by a sum of exponents (see, e.g.,~\cite{Konoplya:2011qq}).

The time-domain integration is an effective method for finding the fundamental mode of asymptotically flat black holes (see \cite{Konoplya:2025hgp,Dubinsky:2024aeu,Qian:2022kaq,Aneesh:2018hlp,Konoplya:2023ppx,Bronnikov:2021liv,Skvortsova:2023zmj,Skvortsova:2024atk,Bronnikov:2019sbx,Malik:2024tuf,Malik:2024sxv,Malik:2024qsz,Jia:2024pdk} for recent examples) and a few first overtones for asymptotically de Sitter spacetimes \cite{Dubinsky:2024gwo,Dubinsky:2024hmn}, as the comparison with other, more accurate methods, shows.

\section{Quasinormal modes of the Reissner-Nordström-de Sitter spacetime}\label{sec:qnms}

\subsection{Oscillatory modes of the charged black holes}\label{sec:complex}

\begin{figure*}
\resizebox{\linewidth}{!}{\includegraphics{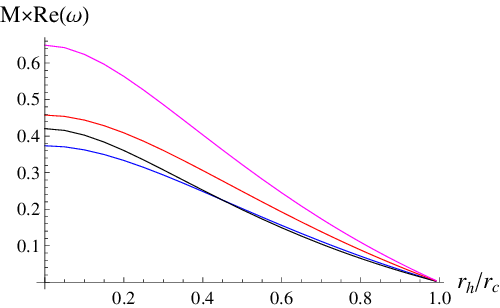}\includegraphics{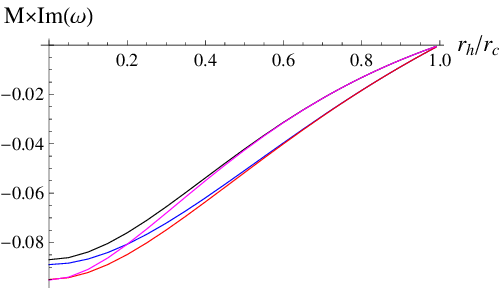}}
\caption{
Dominant $\ell=2$ mode of the black-hole branch (in units of black hole mass) for ``$-$'' (blue) and ``$+$'' (red) perturbations of the uncharged black hole ($r_i = 0$), compared with ``$-$'' (black) and ``$+$'' (magenta) modes of the charged black hole ($r_i = r_h/2$).}
\label{fig:complexmode}
\end{figure*}

The quasinormal modes of the Reissner-Nordström black hole are modified by the presence of a positive cosmological constant. These damped oscillations are characterized by complex frequencies with a nonzero real part. In the asymptotically flat limit, $\Lambda\to0$ (or, equivalently, $r_c\to\infty$), they reduce to the corresponding modes of the Reissner-Nordström solution in asymptotically flat spacetime.

In the opposite limit of large black holes, $r_h\to r_c$, both the real and imaginary parts of the quasinormal frequencies tend to zero \cite{Zhidenko:2003wq}, and the frequencies can be expressed as:
\begin{equation}
\omega=\frac{f'(r_h)}{2}\left(A-\left(n+\frac{1}{2}\right)\imo\right)+\Order{r_c-r_h}^2,
\end{equation}
where $n$ is the overtone number, $A$ is a constant, which depends on the type of perturbations, multipole number $\ell$, and the ratio $r_i/r_h$. For the uncharged black holes \mbox{($r_i=0$)} the value of $A$ matches the corresponding quantity for test fields in the Schwarzschild-de Sitter background \cite{Churilova:2021nnc}, and is given by
\begin{equation}
A=\sqrt{(\ell+s)(\ell+1-s)-\frac{1}{4}},
\end{equation}
where $s=1$ for the electromagnetic perturbations \mbox{(``$+$'' modes)} and $s=2$ for the gravitational perturbations \mbox{(``$-$'' modes)}.

When the electric charge is nonzero ($r_i > 0$), the qualitative behavior of the quasinormal modes in units of the black hole mass remains similar to the charged asymptotically flat case \cite{Leaver:1990zz}. The damping rate changes only slightly with increasing charge, typically decreasing for sufficiently large charge. Meanwhile, the real part of the frequency increases with charge for ``$+$'' modes and also for ``$-$'' modes when the black hole is relatively small.
However, for sufficiently large black holes, specifically when $r_h \gtrsim r_c / 2$, the behavior changes: the ``$-$'' modes of charged black holes may exhibit a smaller real part of the frequency compared to their uncharged counterparts. This nonmonotonic dependence on the charge is illustrated in Fig.~\ref{fig:complexmode}, showing the interplay between the cosmological constant and electric charge in shaping the dominant mode.

\subsection{Zero-damped modes}\label{sec:zdms}
Alongside the oscillatory quasinormal modes, the Reissner-Nordström black hole exhibits a distinct branch that dominates in the near-extremal regime, when the inner horizon approaches the event horizon, $r_h-r_i\ll r_h$. In the nonrotating case, this family has a vanishing real part and therefore consists of purely imaginary frequencies, corresponding to nonoscillatory, exponentially decaying modes. In the literature on charged black holes, these modes have been referred to as quasibound states \cite{Kim:2012mh} and near-extremal modes \cite{Cardoso:2017soq}.

Subsequent work clarified that this family is not distinct from the zero-damped modes originally identified for Kerr black holes \cite{Detweiler:1980gk}. In particular, it was shown in \cite{Zimmerman:2015trm} that, for Kerr-Newman spacetimes, these modes belong to the same branch as the Kerr ZDMs. In the rotating case, the modes acquire a nonvanishing real part proportional to the horizon’s frame-dragging angular velocity, whereas in the nonrotating limit this real part vanishes, reducing the spectrum to purely imaginary frequencies. ZDMs scale with the surface gravity of the inner horizon as \cite{Zimmerman:2015trm}
\begin{equation}\label{ZDMGEM}
\omega=\imo f'(r_i)\frac{\ell+n+1+(-1)^{s-1}}{2}+\Order{\frac{r_h-r_i}{r_h}}^2,
\end{equation}
where $n=0,1,2,\ldots$, $s=1$ for \mbox{``$+$'' modes} and $s=2$ for \mbox{``$-$'' modes}. We have verified that for the $\ell=1$ pure Maxwell perturbations (\ref{l1Maxwell}), the ZDMs approach
\begin{equation}
\omega=\imo f'(r_i)\frac{n+3}{2}+\Order{\frac{r_h-r_i}{r_h}}^2,
\end{equation}
coinciding with (\ref{ZDMGEM}) for $\ell=s=1$.

\begin{figure}
\resizebox{\linewidth}{!}{\includegraphics{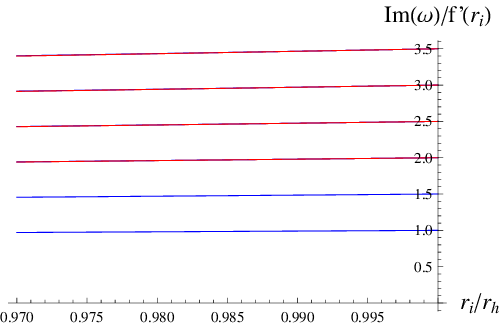}}
\caption{
Lowest zero-damped modes of the near extreme Reissner-Nordström black hole ($\Lambda=0$) for $\ell =2$, in units of $f'(r_i)<0$. The blue and red curves correspond to the ``$-$'' and ``$+$'' perturbation sectors, respectively. Within the ``$-$'' perturbation sector, the two lowest modes are the dominant ones. The ``$+$'' sector contains the modes with larger imaginary parts, which coincide with the higher-damped ``$-$'' modes in the near-extremal limit $r_i\to r_h$.}
\label{fig:ZDMr0}
\end{figure}

By fitting the numerical data in the regime $r_h-r_i\ll r_h$, we find that the leading correction to the spectrum of ZDMs can be written as
\begin{eqnarray}
\omega&=&\imo f'(r_i)\frac{\ell+n+1+(-1)^{s-1}}{2}\\\nonumber&&\times\left(1+B\frac{r_h-r_i}{r_h}+\Order{\frac{r_h-r_i}{r_h}}^2\right),
\end{eqnarray}
where the coefficient $B$ is universal: it does not depend on $\ell$, $n$, or $s$, and is numerically close to $B\approx-1$. Higher-order corrections of order $(1-r_i/r_h)^2$ depend strongly on $\ell$ and $n$ and also differ between the $+$ and $-$ sectors.
However, in the near-extreme regime these contributions are negligible, and ZDMs of the ``$+$'' ($s=1$) channel coincide with the $n\ge2$ ZDMs of the ``$-$'' ($s=2$) channel, as shown in Fig.~\ref{fig:ZDMr0}.

\begin{figure}
\resizebox{\linewidth}{!}{\includegraphics{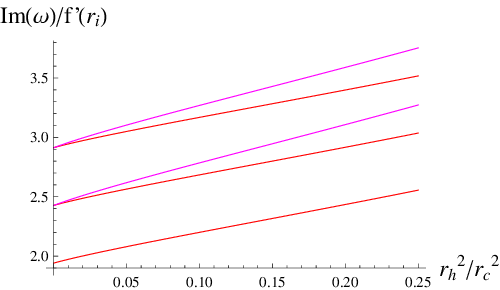}}
\caption{
Lowest zero-damped modes of the near extreme Reissner-Nordström-de Sitter black hole with $r_i=0.97r_h$, for ``$+$'' perturbations, in units of $f'(r_i)<0$: $\ell =2$ (red, lower slope) and $\ell =3$ (magenta, higher slope).}
\label{fig:ZDMrho}
\end{figure}

\begin{figure}
\resizebox{\linewidth}{!}{\includegraphics{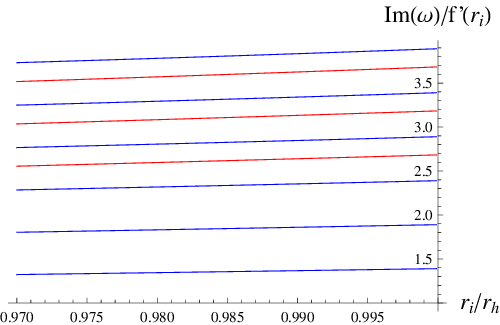}}
\caption{
The nine lowest zero-damped modes of the near-extremal Reissner-Nordström-de Sitter black hole with \mbox{$r_c=2r_h$}, for $\ell =2$, in units of $f'(r_i)<0$. The blue and red curves correspond to the ``$-$'' and ``$+$'' perturbation sectors, respectively. From bottom to top, the first (dominant) three modes belong to the ``$-$'' sector, after which the branches alternate between the ``$+$'' and ``$-$'' sectors.}
\label{fig:ZDMr05}
\end{figure}

For a nonzero cosmological constant, zero-damped modes persist and remain purely imaginary. They again approach zero as $r_i\to r_h$, reflecting the vanishing surface gravity of the degenerate horizon. Fitting numerical data for $r_h\ll r_c$ in the regime $r_h-r_i\ll r_h$, we find
\begin{eqnarray}\label{ZDMdS}
\omega&=&\imo\frac{f'(r_i)}{2}\left(\ell+n+1+(-1)^{s-1}\right)\\\nonumber&&\times\left(\frac{r_i}{r_h}+C\frac{r_h}{r_c}+\Order{\frac{r_h-r_i}{r_h}}^2+\Order{\frac{r_h}{r_c}}^2\right),
\end{eqnarray}
where $C\approx0$ for all $\ell$, $n$, $s$, and $r_i$.
Thus the leading cosmological-constant correction enters only at order $r_h^2/r_c^2$ (see Fig.~\ref{fig:ZDMrho}), and the slope depends on $\ell$, $n$, $s$, and also $r_i$. In particular, when the black-hole and cosmological horizons become comparable, the spectra of zero-damped modes for ``$+$'' ($s=1$) and ``$-$'' ($s=2$) channels of the near-extremely charged black holes differ significantly, as seen in Fig.~\ref{fig:ZDMr05}.

Finally, we note that when the $\Lambda$-term is small ($r_h\ll r_c$) the derivative of the metric function at the inner horizon behaves as
$$f'(r_i)=-\frac{r_h-r_i}{r_i^2}+\Order{\frac{1}{r_c}}^2,$$
and our formula (\ref{ZDMdS}) takes a simple explicit form in terms of the geometric units
\begin{eqnarray}\nonumber
\omega&=&\imo f'(r_i)\frac{\ell+n+1+(-1)^{s-1}}{2}\\\nonumber&&\times\left(\frac{r_i}{r_h}+\Order{\frac{r_h-r_i}{r_h}}^2+\Order{\frac{r_h}{r_c}}^2\right)\\
\label{analyticZDM}
&=&-\imo(r_h-r_i)\frac{\ell+n+1+(-1)^{s-1}}{2r_hr_i}\\\nonumber&&\times\left(1+\Order{\frac{r_h-r_i}{r_h}}^2+\Order{\frac{r_h}{r_c}}^2\right).
\end{eqnarray}
This explicitly shows that zero-damped modes of the Reissner-Nordström-de Sitter black holes are controlled by the near-horizon geometry of the highly charged black hole with only subleading corrections from the cosmological scale.

\subsection{de Sitter modes}\label{sec:deSitter}

In addition to the quasinormal modes associated with the black hole, which approach the Reissner-Nordström spectrum in the limit \mbox{$r_c\to\infty$}, there exists another branch of purely imaginary quasinormal modes. As shown for the test scalar field \cite{Cardoso:2017soq} and later for gravitational perturbations \cite{Konoplya:2022xid}, when \mbox{$r_h\to 0$}, these purely imaginary modes go over into the modes of empty de Sitter spacetime, $\omega_{n}^{(dS)}$, which have the following form~\cite{Lopez-Ortega:2006tjo,Lopez-Ortega:2006aal}:
\begin{equation}\label{dSmodes}
\omega_{n}^{(dS)}r_{c} = -\imo (\ell + n+1-\delta_{s0}\delta_{n0}),
\end{equation}
where $n=0,1,2,\ldots$ is the overtone number and $\ell=s,s+1,\ldots$ is the multiple number.

The presence of purely imaginary de Sitter modes in the quasinormal-mode spectrum of black holes is particularly interesting, because such modes do not obey the well-known correspondence between eikonal quasinormal modes and null geodesics. According to this correspondence, valid in the limit $\ell\to\infty$, the real and imaginary parts of the quasinormal frequencies are related to the angular frequency and the Lyapunov exponent of the unstable null geodesics, respectively~\cite{Cardoso:2008bp}. This relation holds due to the dominance of the centrifugal term $\propto f(r)\ell^2/r^2$ in the effective potential. Consequently, when the centrifugal term takes an unusual form -- as it does in various higher-curvature gravity theories -- the correspondence no longer holds~\cite{Konoplya:2025afm,Konoplya:2017wot,Konoplya:2020bxa}. The correspondence between null geodesics and the eikonal behavior of quasinormal modes also breaks down for electromagnetic perturbations in nonlinear electrodynamics, where light rays do not follow null geodesics~\cite{Toshmatov:2019gxg}. In our case, although the theory involves standard Maxwell electrodynamics and the centrifugal term retains its usual form, the correspondence still breaks down because the de Sitter branch of modes cannot be captured by the WKB approximation, which is essential for establishing this correspondence~\cite{Konoplya:2022gjp}.
The black-hole branch alone still obeys the correspondence. In addition, the de Sitter branch determines the strong cosmic censorship bound for quasinormal modes when the radius of the event horizon is much smaller than the de Sitter radius~\cite{Konoplya:2022kld}.

If the radius of the black hole is much smaller than the radius of the cosmological horizon, the quasinormal modes of test fields obey the following {\it universal law} \cite{Konoplya:2022kld}, which is valid not only for the Schwarzschild-de Sitter black hole, but also for any spherically symmetric black hole independently on an underlying metric gravitational theory:
\begin{equation}\label{BHmodes}
\omega_n = \omega_{n}^{(dS)}\left(1-\frac{M}{r_{c}}+\Order{\frac{M}{r_{c}}}^2\right).
\end{equation}

\begin{figure}
\resizebox{\linewidth}{!}{\includegraphics{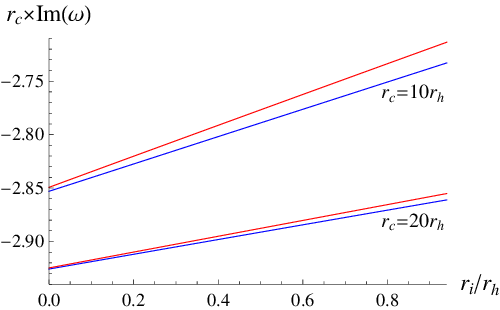}}
\caption{
Lowest purely imaginary quasinormal modes for $\ell=2$ as functions of the charge parameter $r_i/r_h$. Blue (lower) and red (upper) curves correspond to the ``$-$''
and ``$+$'' perturbation sectors, respectively.}
\label{fig:ri}
\end{figure}

For the coupled electromagnetic and gravitational perturbations the dependence of the purely imaginary branch of modes is remarkably close to the linear function of $r_i$ (see Fig.~\ref{fig:ri}).

By fitting the numerical data for $r_h\ll r_c$ across various multipoles and overtones, we obtain the following approximate formula:
\begin{equation}\label{ridependence}
\omega_n \propto 1-\frac{r_i}{2r_c},
\end{equation}
which confirms that the universal law (\ref{BHmodes}) holds for the gravitational and electromagnetic perturbations of the charged black holes,
$$\frac{\omega_n}{\omega_{n}^{(dS)}}=1-\frac{M}{r_{c}}+\Order{\frac{M}{r_{c}}}^2 \!\!\! = 1-\frac{r_h+r_i}{2r_{c}}+\Order{\frac{r_h}{r_{c}}}^2 \!\!\! .$$

\begin{figure}
\resizebox{\linewidth}{!}{\includegraphics{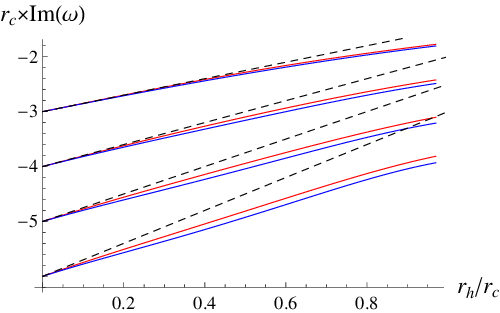}}
\caption{
Lowest purely imaginary $\ell=2$ quasinormal modes of the uncharged Reissner-Nordström-de Sitter black hole ($r_i=0$).
Blue (lower) and red (upper) curves correspond to the ``$-$'' and ``$+$'' perturbation sectors, respectively. The black dashed curves show the universal small-black-hole approximation~(\ref{BHmodes}). The agreement improves as $r_h/r_c\to0$, where these modes approach the purely imaginary de Sitter branch.}
\label{fig:ri0}
\end{figure}

Surprisingly, the above approximation remains accurate for describing the de Sitter branch of modes even in the presence of a large black hole. As the black hole mass increases, the decay rate of the purely imaginary mode decreases. However, as illustrated in Fig.~\ref{fig:ri0}, it does not vanish in the near-extremal limit, since the black hole mass satisfies $M < r_h \to r_c$ and never exactly reaches the cosmological horizon.

\begin{figure*}
\resizebox{\linewidth}{!}{\includegraphics{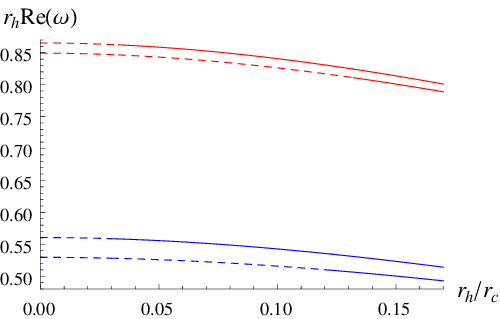}\includegraphics{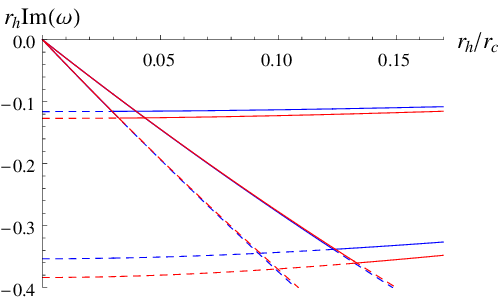}}
\caption{
Two least-damped $\ell=2$ modes of a charged black hole with $r_i = r_h/2$, shown for ``$-$'' (blue) and ``$+$'' (red) perturbations. Solid lines indicate the least-damped modes, while dashed lines represent the corresponding modes in the parametric region, when they are not the least damped. For small black holes, the least damped modes are purely imaginary.}
\label{fig:dominantmodes}
\end{figure*}

\subsection{Interplay between the branches of the quasinormal modes}\label{sec:dominantmodes}

We are now in a position to present a complete picture of the quasinormal spectrum in Reissner-Nordström-de Sitter spacetime. It consists of three distinct branches, one of which contains oscillatory modes with a nonzero real part of the frequency, while the other two are nonoscillatory:
\begin{enumerate}
\item \emph{The oscillatory quasinormal modes} that approach the Reissner-Nordström spectrum in the asymptotically flat spacetime limit. These modes become the least damped for large black holes, and in the near-extreme limit ($r_h \to r_c$), their frequencies scale with $r_c-r_h$.
\item \emph{The de Sitter branch}, consisting of purely imaginary modes, which are the least damped when the black hole is small ($r_c \gg r_h$), because the frequencies scale roughly as $r_c^{-1}$.
\item \emph{Zero-damped modes} for the near-extremally charged black holes, which are also purely imaginary and become the least damped for the sufficiently large electric charge, because the frequencies scale as $r_h-r_i$.
\end{enumerate}

The first two branches lead to a peculiar and nonmonotonic behavior of the least damped mode as a function of the black hole radius $r_h$. For small black holes, the purely imaginary modes of the de Sitter branch have the slowest decay rate. As $r_h$ increases, the damping rate of this mode increases, and eventually a mode of the black-hole branch becomes the least damped. Upon further increase of $r_h$, the next mode of the black-hole branch (the first overtone) overtakes the purely imaginary mode. As the black hole approaches its extremal size ($r_h \to r_c$), more and more of the dominant overtones transition from purely imaginary to oscillatory modes, while the purely imaginary modes are pushed to higher overtone numbers.

Fig.~\ref{fig:dominantmodes} illustrates this transition for charged black holes of varying sizes. For sufficiently small black holes, the two least-damped modes are purely imaginary. At a certain value of $r_h$, the least-damped $\ell=2$ ``$-$'' mode becomes one from the black-hole branch, followed by a similar transition for the $\ell=2$ ``$+$'' mode. After that, the least-damped purely imaginary mode becomes the first overtone. At larger $r_h$, the second black-hole mode becomes the first overtone -- first for ``$-$'' perturbations, and later for ``$+$''. The higher the ratio $r_h/r_c$, the more overtones transition to modes from the black-hole branch.

It is important to note that, although the purely imaginary modes are the least damped in Reissner-Nordström-de Sitter spacetime when $r_h \ll r_c$, they are typically not observed in the gravitational-wave signal emitted by a perturbed black hole. These modes appear only at very late times as the so-called exponential asymptotic tails, when the gravitational-wave signal has already decayed significantly. The reason is that the initial perturbation tends to excite the modes associated with the black hole much more strongly than the modes associated with the de Sitter Universe. Therefore, one must carefully distinguish between the full quasinormal spectrum of a spacetime and the subset of modes that are actually observable in a gravitational-wave signal.

Similarly, when the black hole charge approaches its extremal value, ZDMs become the least damped, shifting from higher overtones to the dominant nonoscillating modes in the ringdown. Unlike de Sitter modes, they are associated with the black hole horizon, reflecting the existence of a hidden conformal symmetry in the near-horizon geometry \cite{Kim:2012mh,Hintz:2025nij}. As a result, when a black hole is perturbed, ZDMs are excited along with the oscillatory quasinormal modes and, for near-extremal black holes, can be observed in the time domain \cite{Konoplya:2013rxa}.

\begin{figure*}
\resizebox{\linewidth}{!}{\includegraphics{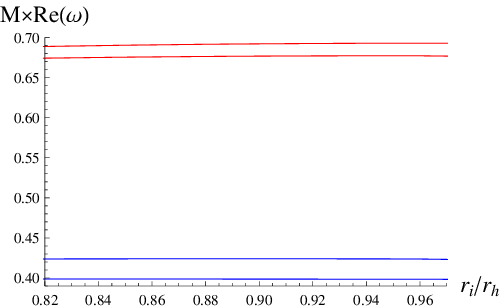}\includegraphics{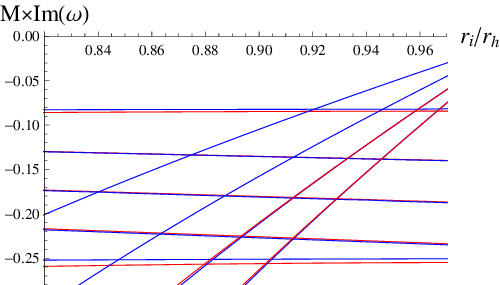}}
\caption{
Real part (left panel) and imaginary part (right panel) of the least-damped $\ell=2$ modes of a near-extremally charged Reissner-Nordström-de Sitter black hole with $r_c=20r_h$ ($\Lambda M^2\simeq0.002$). The blue and red curves correspond to the ``$-$'' and ``$+$'' perturbation sectors, respectively, and the frequencies are shown in units of the black-hole mass. As the inner horizon approaches the event horizon, the purely imaginary zero-damped modes become dominant. In this regime the oscillatory modes in the two sectors have nearly the same decay rate, although the ``$-$'' mode is slightly longer-lived and has a substantially smaller oscillation frequency. The next oscillatory mode shown is not the first overtone: three purely imaginary de Sitter modes have smaller damping rates, which increase as the black-hole charge approaches its extremal value. By contrast, the damping rates of the purely imaginary zero-damped modes rapidly decrease near extremality, causing these modes to become dominant. For clarity, only six zero-damped modes are displayed: four in the ``$-$'' sector and two in the ``$+$'' sector, the latter lying close to the corresponding ``$-$'' branches. The two least damped ZDMs in the ``$-$'' sector decay substantially more slowly than their ``$+$'' counterparts. Consequently, they overtake the oscillatory modes and become dominant at smaller values of the charge.}
\label{fig:nearextreme}
\end{figure*}

\begin{figure*}
\resizebox{\linewidth}{!}{\includegraphics{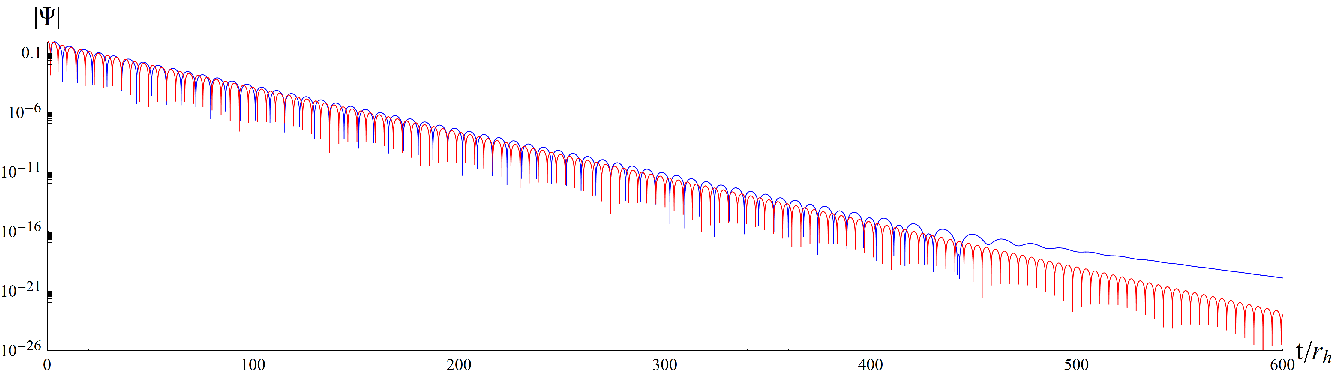}}
\caption{
Time-domain profiles of the $\ell=2$ ``$-$'' (blue) and ``$+$'' (red) perturbations of the Reissner-Nordström-de Sitter black hole for $r_c=20r_h$ and $r_i=0.95r_h$. For this near-extremal value of the electric charge, the zero-damped modes are already dominant in the ``$-$'' sector, producing a nonoscillatory late-time decay. In the ``$+$'' sector the dominant contribution is still given by an oscillatory mode, and the ringdown therefore remains oscillatory for all time.}
\label{fig:timedomain}
\end{figure*}

The dominant modes of near-extremally charged Reissner-Nordström-de Sitter black holes are shown in Fig.~\ref{fig:nearextreme}. As follows from Eq.~(\ref{analyticZDM}), the least-damped ZDMs have a smaller decay
rate in the ``$-$'' perturbation sector. For $r_c=20r_h$, these modes become dominant in the ``$-$'' sector for $r_i\gtrsim0.92r_h$, whereas in the ``$+$'' sector they become dominant only for $r_i\gtrsim0.96r_h$.

When zero-damped modes dominate the spectrum, the late-time signal becomes nonoscillatory at late times. This behavior is illustrated in Fig.~\ref{fig:timedomain}, where we show the time-domain profiles in both perturbation sectors for $r_i=0.95r_h$. At this value of the electric charge, the ZDMs dominate only in the ``$-$'' sector, while the ``$+$'' sector is still dominated by an oscillatory ringdown mode.

\begin{table}
\begin{tabular}{l|c|c}
\hline
Mode & Frequency domain & Time domain \\
\hline
``$-$'' ZDM & $~~~~~~~~~~~~-0.052278\imo$ & $~~~~~~~~~~-0.05228\imo$\\
``$-$'' ZDM & $~~~~~~~~~~~~-0.078374\imo$ & $~~~~~~~~~~-0.07870\imo$\\
\hline
``$-$'' oscillatory & $0.436541-0.084671\imo$ & $0.43658-0.08466\imo$\\
``$-$'' oscillatory & $0.410390-0.258288\imo$ & $0.41098-0.25827\imo$\\
\hline
``$+$'' oscillatory & $0.713841-0.086914\imo$ & $0.71384-0.08691\imo$\\
``$+$'' oscillatory & $0.697765-0.262694\imo$ & $0.69797-0.26261\imo$\\
\hline
\end{tabular}
\caption{
Dominant $\ell=2$ modes, in units of the horizon radius $r_h$, for $r_c=20r_h$ and $r_i=0.95r_h$. The frequencies computed in the frequency domain by the accurate continued-fraction method are compared with the values extracted from the time-domain profiles shown in Fig.~\ref{fig:timedomain}. In the ``$-$'' sector, the least-damped modes are purely imaginary ZDMs and hence produce a nonoscillatory late-time decay. In the ``$+$'' sector, the dominant modes are still oscillatory, so that the corresponding time-domain ringdown profile remains oscillatory for all time.}
\label{table}
\end{table}

By fitting the late-time profile one can extract the zero-damped modes in the ``$-$'' perturbation sector. The subdominant oscillatory modes are instead most accurately obtained from the earlier, oscillatory stage of the ringdown. The results are summarized in Table~\ref{table}: ZDMs were extracted from the nonoscillatory late-time regime, while the oscillatory modes were extracted from the initial ringdown phase.

As the black-hole charge is increased further, additional ZDMs become dominant. Eventually, the ``$+$'' perturbation sector also becomes nonoscillatory at late times. A similar effect occurs when $r_c$ is increased,
or equivalently when $\Lambda$ is decreased: the purely imaginary de Sitter modes become dominant, leading to nonoscillatory asymptotic behavior in both perturbation sectors.

\section{Conclusions}\label{sec:conclusions}

The Reissner-Nordström-de Sitter metric describes a classical black hole solution to the Einstein-Maxwell equations with a cosmological constant. As a result, perturbations and quasinormal modes of both test fields and gravitational perturbations have been extensively studied in the literature \cite{Zhu:2014sya,Zhang:2020zic,Fontana:2020syy,Nan:2023vkq,Chrysostomou:2023jiv,Chrysostomou:2025twu,Liu:2024eut}. However, the spectrum of gravitational perturbations has not been fully explored to date. Of the black-hole and de Sitter modes, only the former has been studied in detail \cite{Molina:2003dc}. In this work, we address this gap by investigating the previously unexplored de Sitter branch, thereby providing a more complete picture of the quasinormal mode spectrum of this classical black hole solution.
We conducted a comprehensive analysis of the full quasinormal spectrum of the coupled gravitational and electromagnetic perturbations of the Reissner-Nordström-de Sitter black hole. Our study reveals that purely imaginary modes of the de Sitter branch emerge within the gravitational perturbation spectrum of charged black holes and satisfy the universal analytic formula in the regime of small black holes, which was previously obtained only for test fields. This analytic expression provides a good approximation for the modes of the de Sitter branch even as the black hole size is comparable with the de Sitter radius. In addition, we studied the zero-damped modes of Reissner-Nordström-de Sitter perturbations and computed numerically the higher-order corrections to the analytic formula for near-extremal charges.

Although the quasinormal modes of the de Sitter branch are the least damped for small black hole sizes, these modes are not typically observed in gravitational-wave signals. This is because the modes from the black-hole branch tend to be the most strongly excited by the perturbations of black holes. As a result, the modes of the de Sitter branch primarily manifest only as late-time exponential tails, appearing after the main signal has significantly decayed. Consequently, their observational signatures in gravitational-wave data are generally suppressed and difficult to detect.

\bibliography{bibliography}
\end{document}